\documentclass[twocolumn,prl,superscriptaddress,showpacs]{revtex4}
\usepackage{epsf,graphicx,amssymb}
\begin{document}
\draft
\title{Bypassing Cowling's theorem in axisymmetric fluid dynamos}

\author{Christophe Gissinger}
\affiliation{Laboratoire de Physique Statistique de l'Ecole Normale Sup\'erieure, CNRS UMR 8550, 24 Rue Lhomond, 75231 Paris Cedex 05, France}
\author{Emmanuel Dormy}
\affiliation{MAG (IPGP/ENS), CNRS UMR 7154, LRA, Ecole Normale Sup\'erieure, 24 Rue Lhomond, 75231 Paris Cedex 05, France}
\author{Stephan Fauve}
\affiliation{Laboratoire de Physique Statistique de l'Ecole Normale
Sup\'erieure, CNRS UMR 8550, 24 Rue Lhomond, 75231 Paris Cedex 05, France}

\def\bfnabla{\mbox{\boldmath $\nabla$}}
\def\bfv{\mbox{\boldmath $v$}}
\def\bfF{\mbox{\boldmath $F$}}
\def\bff{\mbox{\boldmath $f$}}

\date{\today}
\begin{abstract}
We present a numerical study of the magnetic field generated by an
axisymmetrically forced flow in a spherical domain. At small enough
Reynolds number, $Re$, the flow is axisymmetric and generates an
equatorial dipole above a critical magnetic Reynolds number
$Rm_{c}$. The magnetic field thus breaks axisymmetry, in agreement
with Cowling's theorem. This structure of the magnetic field is
however replaced by a dominant axial dipole when $Re$ is larger 
and allows non axisymmetric fluctuations in the flow. 
We show here that even in the absence of such fluctuations,
an axial dipole can also be generated, at low $Re$,
through a secondary bifurcation, when $Rm$ is increased above the
dynamo threshold. The system therefore always find a way to bypass the
constraint imposed by Cowling's theorem. We understand the dynamical
behaviors that result from the interaction of equatorial and axial
dipolar modes using simple model equations for their amplitudes 
derived from symmetry arguments.  
\end{abstract}
\maketitle 
 
It is strongly believed that magnetic fields of planets and stars are
generated by dynamo action, i.e., self generation of a magnetic field by
the flow of an electrically conducting fluid~\cite{moffatt}.  Planets
and stars being rapidly rotating, axisymmetric flows about the axis of
rotation have often been considered in order to work out simple dynamo
models~\cite{dudley}. A major setback of the subject followed the
discovery of Cowling's theorem, which stated that a purely magnetic field cannot be maintained by dynamo
action~\cite{cowling}.
However, it has been shown that magnetic fields with a dominant
axisymmetric mean part can be generated when non-axisymmetric
helical fluctuations are superimposed to a mean axisymmetric flow
\cite{alpha}. This  has been recently observed in the VKS
experiment~\cite{VKS}. A strongly turbulent swirling von
K\'arm\'an flow driven by two counter-rotating coaxial disks in a
cylindrical container self-generated a magnetic field with a
dipole mean component along the axis of rotation. This has been
ascribed to an alpha effect due to the helical nature of the
radially ejected flow along the two impellers 
\cite{petrelis07}. In this letter, we show that there exists
another mechanism for bypassing the constraint imposed by
Cowling's theorem, without the help of non axisymmetric turbulent
fluctuations. The mechanism is as follows: the primary dynamo
bifurcation breaks axisymmetry in agreement with Cowling's
theorem. Then, the Lorentz force generates a non axisymmetric flow
component which can drive an axisymmetric magnetic field through a
secondary bifuraction. We show that direct numerical simulations
confirm this scenario and that the two successive bifurcation
thresholds can be very close in some flow configurations. The
existence of two competing instability modes, the axial and
equatorial dipoles, can lead to complex dynamical 
behaviors. Using symmetry arguments, we write equations for the 
amplitude of these modes that are coupled through the non
axisymmetric velocity component. We show that the observed
bifurcation structure and the resulting dynamics can be understood
in the framework of this simple model.\\
We first numerically integrate the MHD equations in a spherical
geometry for the solenoidal velocity $\bf v$ and magnetic $\bf B$
fields,
\begin{eqnarray}
\frac{\partial \bf{v} }{ \partial t} + (\bf{v}\cdot\bfnabla) \bf{v} \!&=&\! -
\bfnabla \pi + \nu 
\Delta \bf{v} + \bf{f} +\frac{1}{\mu \rho}(\bf {B} \cdot \bfnabla) \bf
{B}, \label{ns}\\
\frac{\partial \bf{B}}{ \partial t} \!&=&\! \bfnabla
\times \left(\bf{v} \times \bf{B}\right) + \eta \, 
\Delta \bf {B}. \label{ind}
\end{eqnarray}
In the above equations, $\rho$ is the density, $\mu$ is the magnetic
permeability and $\sigma$ is the conductivity of the fluid. 
The forcing is ${\bf f} = f_0\,{\bf F}$, where
$F_\phi=s^2 \sin(\pi\, s \,b)\, , \; F_z=\varepsilon\, \sin(\pi \, s \,
c) \, ,$ for $z>0$, using polar coordinates $(s, \phi, z)$ (normalized
by the radius of the sphere $a$) and opposite for $z<0$.
$F_\phi$ generates counter-rotating flows in each hemisphere, while
$F_z$ enforces a strong poloidal circulation. The forcing is only
applied in the region $0.25a<\mid\!z\!\mid<0.65a$, $s<s_0$.  In the
simulations presented here, $s_0=0.4$, $b^{-1}=2s_0$ and
$c^{-1}=s_0$.  This forcing has previously been introduced to
model the mechanical forcing due to co-axial rotating impellers used
in the Madison experiment~\cite{bayliss07}.  Although performed in a
spherical geometry, this experiment involves a mean flow with a
similar topology to that of the VKS experiment. Such flows correspond
to $s_2+t_2$ flows in the Dudley and James classification
\cite{dudley}, i.e. two poloidal eddies with inward flow in the
mid-plane, together with two counter-rotating toroidal eddies.
We solve the above system of equations using the Parody numerical code
\cite{parody}. This code was originally developped in the context of
the geodynamo (spherical shell) and we have here modified the code to
make it suitable for a full sphere.
We use the same dimensionless numbers as in \cite{bayliss07}, 
the magnetic Reynolds number
$Rm= \mu_0\sigma a \, {\rm max} (\vert \bf v \vert) $, and
the magnetic Prandtl number $Pm=\nu\mu_0\sigma$.
The kinetic Reynolds number is then 
$Re=Rm/Pm$.
\begin{figure}
\centerline{
\epsfxsize=0.3\textwidth 
\epsffile{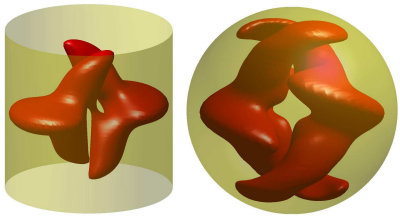} }
\vskip -4mm
\caption{Comparison of the magnetic field generated by an axisymmetric
  $s2+t2$ flow in different geometries. Isovalue of the magnetic energy in a
  cylinder (from \cite{Kris1}) or a sphere.}
\label{banana}
\end{figure}
\begin{figure}
\centerline{
\epsfxsize=0.45\textwidth 
\epsffile{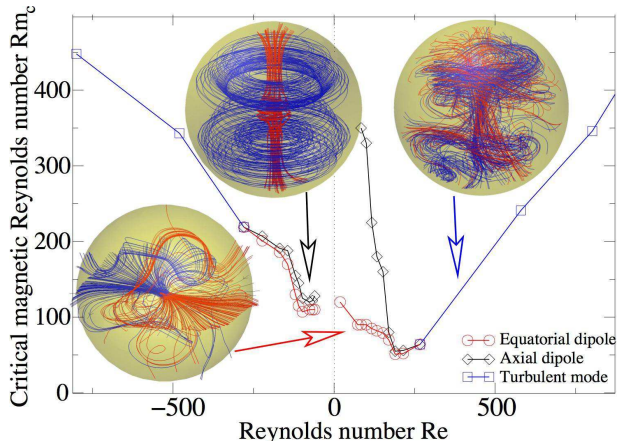} }
\vskip -4mm
\caption{
Stability curves $Rm_c=f(Re)$ obtained with direct numerical
  simulations. In red: onset of the $m=1$ (equatorial dipole) dynamo mode;
  black: non-linear threshold of the $m=0$ (axisymmetric) mode; in blue
  the turbulent mode emerging from velocity fluctuations. The
  corresponding magnetic structures are represented using magnetic
  field lines.}
\label{stabcurve}
\end{figure}

The dynamo threshold $Rm_{c}$ is displayed as a function of $Re$ in
Fig.~\ref{stabcurve}. Negative $Re$ corresponds to a flow that is
reversed compared to the VKS configuration, i.e. directed from the
impellers to the center of the flow volume along the axis and radially
outward in the mid-plane. For small enough $Re$, the flow is laminar
and axisymmetric. A magnetic field with a dominant equatorial
dipole mode $m=1$ is generated first (red curve in
Fig.~\ref{stabcurve}). The geometry of the field is displayed in the
left inset of Fig.~\ref{stabcurve} and breaks axisymmetry as expected
from Cowling's theorem. This dynamo mode is similar to that 
obtained in cylindrical geometry, as illustrated in Fig~\ref{banana}.

For $Re$ larger than about $300$, the flow becomes turbulent and the
equatorial dipole is then replaced by a dominant axisymmetric mode
$m=0$. Its threshold increases with $Re$ in the parameter range of the
simulations (blue curve in Fig.~\ref{stabcurve} and its geometry is shown 
in the right inset). These results are in agreement with \cite{bayliss07}. 
It is remarkable
that the axial dipole observed in the VKS experiment and ascribed to non
axisymmetric fluctuations \cite{petrelis07} can also be obtained in
the present simulations even though the level of fluctuations is much
smaller (the parameter range realized in the experiment
being, by far, out of reach of present computer models).

\begin{figure}
\centerline{
\epsfxsize=0.44\textwidth
\epsffile{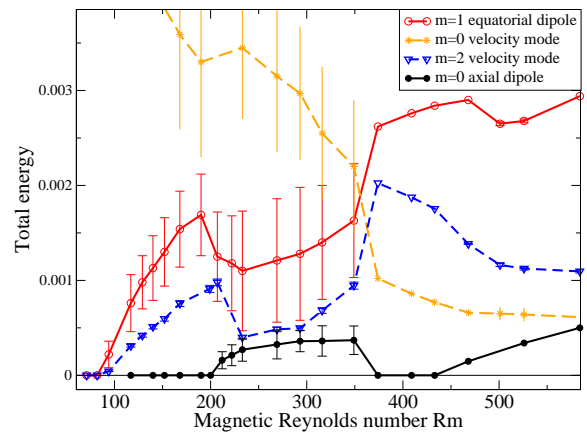} }
\vskip -7mm
\caption{Bifurcation diagram of different magnetic modes varying $Rm$ with fixed $Re =+122$. Error bars indicate the amplitude of oscillations. All other magnetic modes are
very small compared to these ones.}
\label{bifdb}
\end{figure}

In addition, an axisymmetric magnetic field can also be generated at
very low $Re$ through a secondary bifurcation from the equatorial
dipole when $Rm$ is increased.  This corresponds to the black curves
in Fig.~\ref{stabcurve}. The corresponding mode is shown in the top
left inset of Fig.~\ref{stabcurve}. Bifurcation diagram of
Fig.~\ref{bifdb} helps to understand the mechanism by which this
axisymmetric magnetic field is generated. One can observe that the
equatorial dipole first bifurcates supercritically for $Rm = 88$ when
$Re = +122$. The back-reaction of the Lorentz force is twofold. First,
it inhibits the axisymmetric velocity field, which decreases (orange
curve in Fig.~\ref{bifdb}). Second, and more importantly, it drives a
non-axisymmetric $m=2$ velocity mode (blue curve in
Fig.~\ref{bifdb}). Once the intensity of this flow becomes strong
enough, it yields a secondary bifurcation of the axisymmetric $m=0$
field mode. This is achieved for $Rm =205 $ (black curve in
Fig. \ref{bifdb}). The amplitude of the equatorial dipole decreases
immediately after this secondary bifurcation.  We observe that the
$m=0$ mode vanishes at higher $Rm$ and then grows again above
$Rm=425$. Although the amplitude of the equatorial and axial
modes behave in a complex manner as $Rm$ is increased, we observe that 
they are anti-correlated, thus showing that they inhibit each other
through the non-linear couplings.

\begin{figure}
\centerline{
\includegraphics[width=.44\textwidth,height=0.24\textheight]{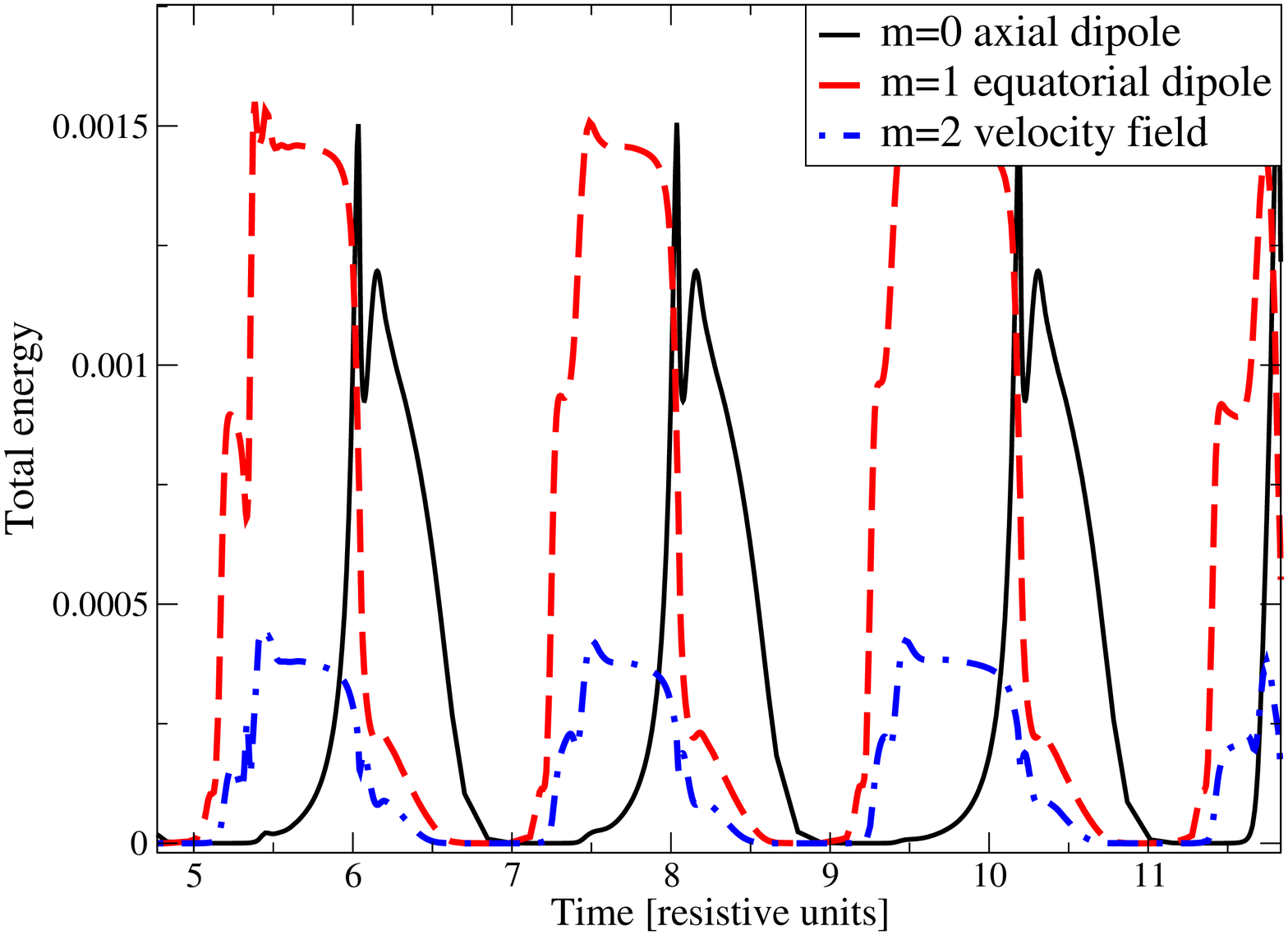}}
\vskip -7mm
\caption{Time recordings of the total energy of the 
equatorial and axial dipolar modes, and of the $m=2$ velocity mode for $Re
 = -76$ and $Rm =170$.} 
\label{time_dns}
\end{figure}

For $Re < 0$, Fig.~\ref{stabcurve} shows that the primary and
secondary bifurcations occur in a much narrower range of $R_m$.  
The equatorial dipole mode is then close to marginal stability
when the axial one bifurcates, and their nonlinear interactions
leads to complex time dependent dynamics close to threshold as 
displayed in Fig.~\ref{time_dns}.    
\begin{figure}
\centerline{
\epsfxsize=0.44\textwidth
\epsffile{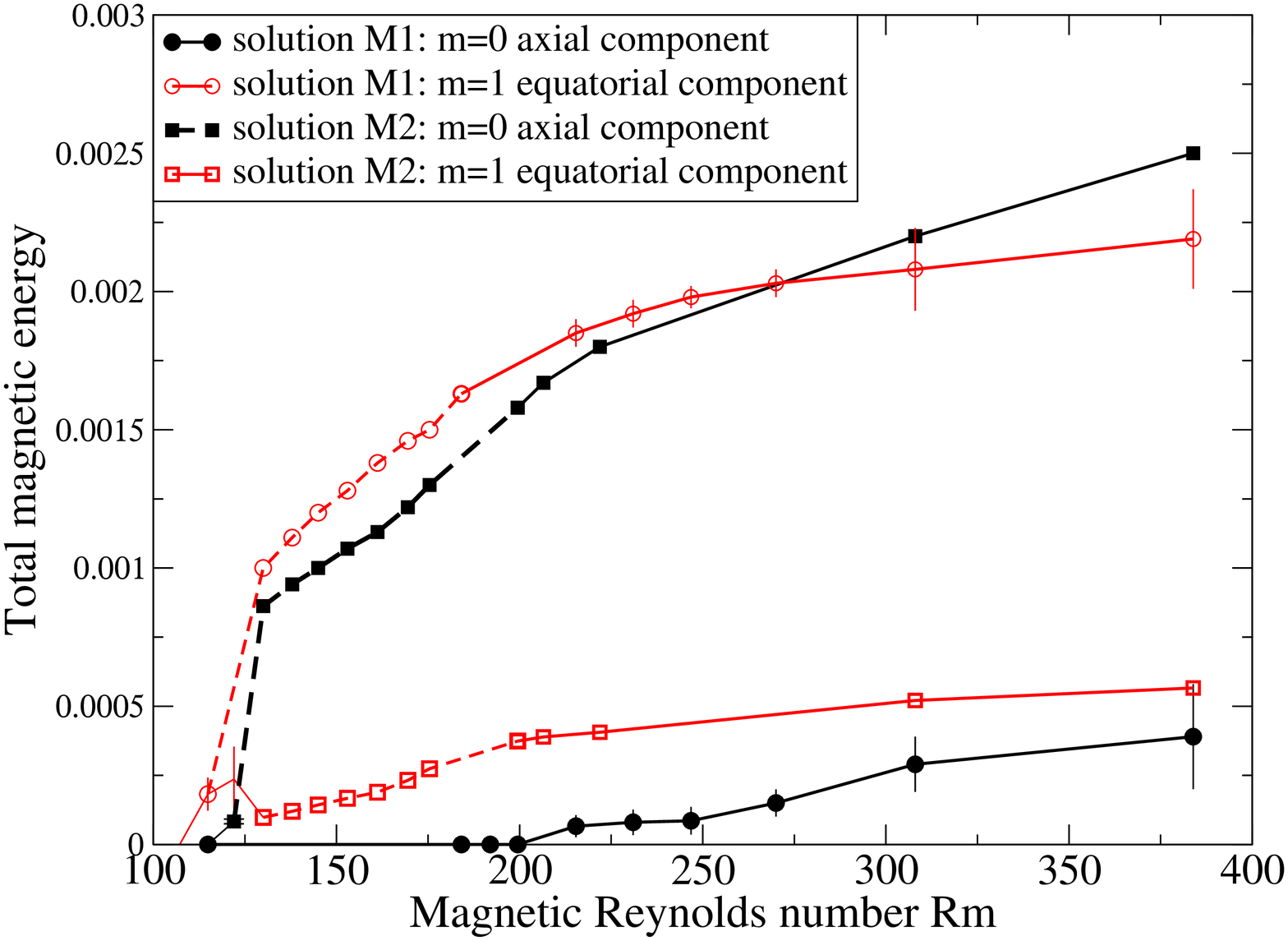} }
\vskip -7mm
\caption{Bifurcation diagram of different magnetic modes varying $Rm$ at fixed $Re =-76$. Error bars indicate the amplitude of oscillations. Dashed lines correspond to the maximum
values of the different modes in the relaxation regime. All other magnetic modes are very weak compared to these ones.}
\label{bifEneg}
\end{figure}
The equatorial mode (red curve) is generated first and saturates, but it
drives the axial mode (black curve) through the non axisymmetric part
of the velocity field.  The axial dipole then inhibits the equatorial
one that decays almost to zero. As a result, the flow is no longer
driven away from axisymmetry by the Lorentz force. The axial
dipole thus decays and the process repeats roughly periodically. 
We observe that during one part of the cycle, the magnetic field is almost
axisymmetric. It involves a strong azimuthal field together with a
large vertical component near the axis of rotation, i.e. an axial
dipole (see the left inset of Fig.~\ref{stabcurve}). These relaxation
oscillations, present only in the $Re<0$ case, occur only slightly
above the threshold of the secondary bifurcation of the $m=0$
mode. Their period first decreases when $Rm$ is increased, but then
increases showing a divergence when the relaxation oscillations
bifurcate to a stationary regime, as displayed in figure~\ref{bifEneg}. 
Above this transition, we observe bistability with   
the coexistence of two solutions: a nearly equatorial dipole, with a
strong equatorial component and a weak axial one (labeled $M_1$ in
figure \ref{bifEneg}) and a nearly axial dipole (labeled $M_2$). 

\begin{figure}
\centerline{
\includegraphics[width=.51\textwidth,height=0.2\textheight]{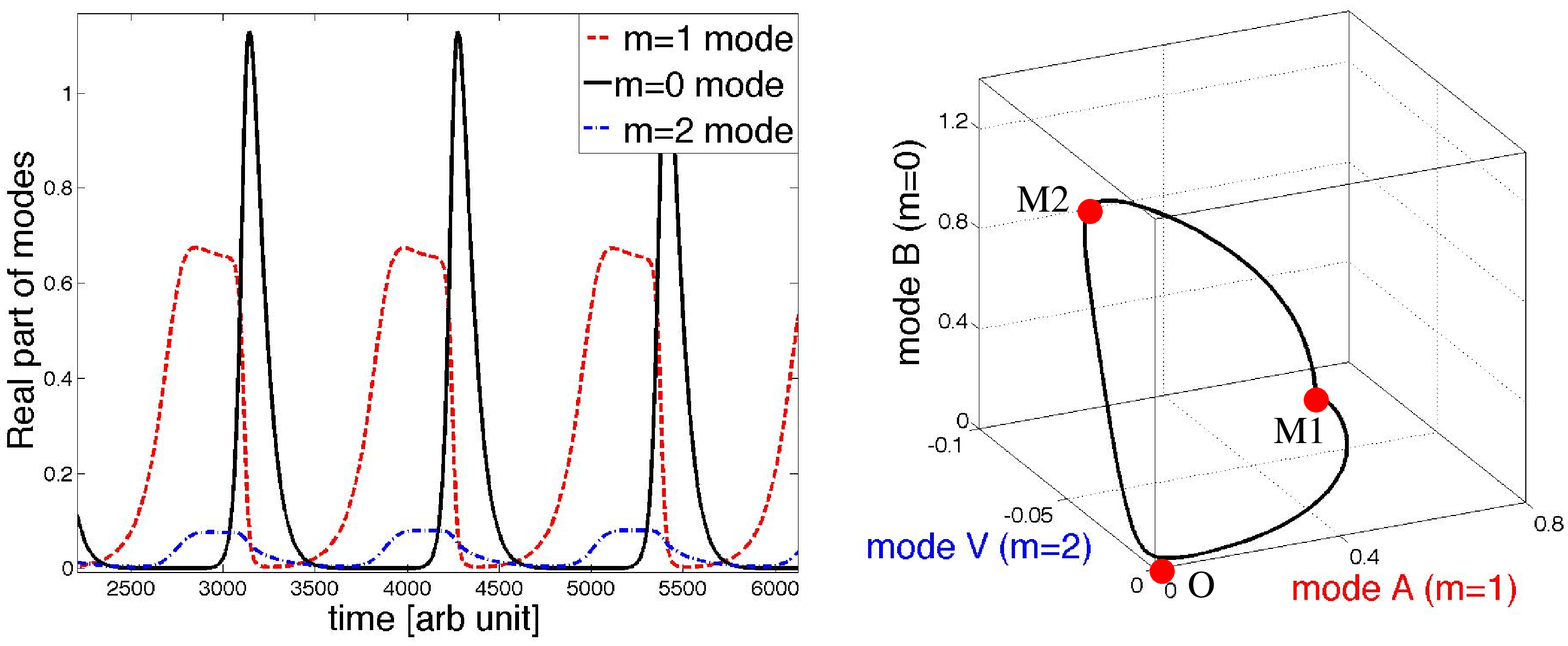}}
\vskip -3mm
\caption{Numerical integration of the amplitude equations
 (\ref{ampA}-\ref{ampV}-\ref{ampB}). Left: time recordings of the
 amplitudes of the equatorial and axial magnetic
 modes interacting through the non axisymmetric velocity mode
 ( $\mu=1,\alpha_1=0.3,\alpha_3=4,\nu=0.5,\beta_2=1.5,\lambda=1.8,\gamma_1=0.5,
\gamma_2=10,\gamma_3=0.5$, all other coefficients being zero.) Right: relaxation cycle in phase space involving the three unstable fixed points (real part of $A$, $V$).}
\label{time_model} 
\end{figure}

We will show next that this competition between equatorial and axial
modes, and the resulting dynamics, can be understood using a
simple model for the amplitudes of the relevant modes.  We thus write
\begin{equation}
{\bf B}({\bf r}, t) = A(t) \, {\bf D}_{eq} ({\bf r}) + c. c. + B(t) \, {\bf D}_{ax} ({\bf r}) + \cdots,
\end{equation}
where ${\bf D}_{eq} ({\bf r})$ (respectively ${\bf D}_{ax} ({\bf r})$) is
the eigenmode related to the equatorial (respectively axial)
dipole. $A$ is a complex amplitude, its phase describes the angle of
the dipole in the equatorial plane and $c. c.$ stands for the complex
conjugate of the previous expression. $B$ is a real amplitude. As said
above, the equatorial dipole ($m=1$) generates a non axisymmetric flow
through the action of the Lorentz force. The later depends
quadratically on the magnetic field, this non axisymmetric
velocity mode of complex amplitude $V(t)$ thus corresponds to $m=2$. Using
symmetry arguments, i.e., rotational invariance about the $z$--axis
which implies the invariance of the amplitude equations under $A \rightarrow
A \exp i\chi, V \rightarrow V \exp 2i\chi$, and the ${\bf B}
\rightarrow -{\bf B}$ symmetry, we get up to the third order
\begin{eqnarray}
\dot{A} &=& \mu A - V \overline{A} - \alpha _1 {\mid\! A \!\mid}^2 A -
\alpha _2 {\mid\! V \!\mid}^2 A - \alpha _3 B^2 A\,, \label{ampA}\\
\dot{V} &=& -\nu V + A^2 - \beta _1 {\mid\! A \!\mid}^2 V - \beta _ 2 {\mid\!
  V \!\mid}^2 V - \beta _3 B^2 V\,, \label{ampV}\\
\dot{B} &=& -\lambda B
-\gamma _1 {\mid\! A \!\mid}^2 B + \gamma _2
{\mid\! V \!\mid}^2 B
-\gamma _3 B^3 \,.
\label{ampB}
\end{eqnarray}
$\mu$ is proportional to the distance to the dynamo threshold. Clearly
$\nu > 0$, since the flow is axisymmetric below threshold. The
coefficients of the quadratic terms can be scaled by an appropriate
choice of the amplitudes. The term $A^2$ represents the forcing of the
non axisymmetric flow by the Lorentz force related to the equatorial
dipole. $V \overline{A}$ means that rotational invariance for the
equatorial dipole is broken as soon as a non axisymmetric flow is
generated. We have fixed its sign so that the bifurcation of the
equatorial dipole remains supercritical $\forall \alpha_1 \geq 0$. The
equations for $A$ and $V$ (with $B=0$) are the normal form of a $1:2$
resonance \cite{resonance1-2} and have been studied in details in
other contexts. In particular, it is known that this system can
undergo a secondary bifurcation for which the phase of $A$ begins to
drift at constant velocity when $\mu$ reaches a value such that
${\mid\!  A \!\mid}^2 = 2 {\mid\! V \!\mid}^2$. 
This corresponds here to a rotating dipole, at constant rate, in the
equatorial plane. Consider now the equation for the amplitude $B$ of
the axial magnetic field. Taking $\lambda > 0$ and $\gamma_3 > 0$
ensures that it cannot be generated alone, in agreement with Cowling's
theorem.  The term ${\mid\! V \!\mid}^2 B$ describes the possible
amplification of $B$ from the non axisymmetric velocity field provided
that $\gamma _2> 0$.  Although the system of amplitude equation
(\ref{ampA}-\ref{ampV}-\ref{ampB}) cannot be derived asymptotically
from (\ref{ns}, \ref{ind}), it reproduces the phenomenology observed with the
direct simulations for both signs of $Re$: when $\mu$ is increased, we
either obtain relaxation oscillations as for $Re<0$ (parameters of
fig.~\ref{time_model}) or a secondary bifurcation of the axial field
as for $Re >0$ (same parameters with $\gamma_2=1$).
The relaxation oscillations are displayed in Fig.~\ref{time_model}
(left). The model helps to understand the qualitative features
observed in the direct simulation: it involves a solution
corresponding to an equatorial dipole ($A_0, V_0, B=0$) that can
bifurcate to a mixed mode $(A_1,V_1, B_1)$ involving a non zero axial
field. In addition, two types of mixed modes can exist, one with a
dominant equatorial dipole, say $M_1 = (A_1, V_1, B_1)$, and another with a
dominant axial dipole $M_2 = (A_2,V_2, B_2)$. Depending on the stability of
these two solutions, we observe either one of the mixed mode
(depending on initial conditions), or a relaxation oscillation slowing
down in the vicinity of these unstable fixed points and the
origin. The system thus has three fixed points with both stable and
unstable directions: the origin where both modes are zero, a point
with a dominant equatorial dipole and a point with a dominant axial
dipole. This situation leads to a heteroclinic cycle connecting these
three unstable equilibrium points and corresponds to the relaxation
oscillations (see fig.~\ref{time_model}, right) .

In sodium flows driven by an axisymmetric forcing, such as
the ones used in the VKS \cite{VKS}, Madison and Maryland experiments
\cite{USboys}, one expects a possible competition between equatorial
and axial dynamo modes. Indeed, the mean flow, if it were acting
alone, would generate an equatorial dipole in agreement with Cowling's
theorem. Our direct simulations show that a fairly small amount of non
axisymmetric fluctuations (compared to the experiments) is enough to
drive an axial ($m=0$) dipole as observed in the VKS experiment for
the mean magnetic field. In addition, we show here that even without turbulent
fluctuations, the non axisymmetric flow driven by the Lorentz force
related to the equatorial dipole, can generate the axial one through a
secondary bifurcation. The equatorial dipole can easily rotate in the equatorial plane, thus averaging to zero. The axial dipole then becomes the dominant part of the mean magnetic field. 

It is striking that this mechanism that generates an axial dipole occurs much closer
to the dynamo threshold when we go from the $Re>0$ to the $Re<0$ flow configuration, thus when the product of the helicity times the differential rotation is changed to its opposite value. For $Re < 0$, the shear layer in the mid-plane becomes favorable to an $\alpha-\omega$ dynamo as soon as the axisymmetry of the flow is broken. For $Re > 0$, the flow near the impellers can play a similar role but the effect is weaker. This opens interesting perspectives for flows that can be used for future dynamo experiments: an $\alpha-\omega$ effect driven by the strong vortices present in the shear layer close to the mid-plane can be favored by the $Re<0$ configuration. To wit, one can use either the optimized set-up described in \cite{petrelis07} or propellors with the appropriate pitch in the VKS or Madison experiments. 

A competition between equatorial and axial dipolar modes could also
account for secular variations of the Earth magnetic field. It would be interesting to check whether some features can be described with a low dimensional model similar to the one used in this study. 

\begin{acknowledgments}
Computations were performed at CEMAG and IDRIS.
\end{acknowledgments}


\end{document}